\documentclass[runningheads,fleqn]{svmult}

\usepackage{makeidx}   
\usepackage{graphicx}  
\usepackage{subeqnar}  
\usepackage{multicol}  
\usepackage{physproc}  
\makeindex             

\begin{document}
\title*{Low-temperature Metastability of Ising Models: 
Prefactors, Divergences, and Discontinuities}
\toctitle{Low-temperature Metastability:
\protect\newline Prefactors, Divergences, and Discontinuities}
%
%
\titlerunning{Low Temperature Metastability:}
%
\author{M.\ A.\ Novotny
}
\authorrunning{Mark A. Novotny}
\institute{Department of Physics and Astronomy\\
and\\
Engineering Research Center\\
P.\ O.\ Box 5167\\
Mississippi State University\\
Mississippi State, MS 39762-5167, U.S.A.}

\maketitle              

\begin{abstract}
The metastable lifetime of the square-lattice and simple-cubic-lattice 
kinetic Ising models are studied in the low-temperature limit.  
The simulations are performed using Monte Carlo with Absorbing Markov 
Chain algorithms to simulate extremely long low-temperature lifetimes.  
The question being addressed is at what temperatures the 
mathematically rigorous low-temperature results become valid.  
It is shown that the answer depends partly on how 
close the system is to fields at which the prefactor for the 
metastable decay either has a discontinuity or diverges.  
\end{abstract}

\section{Introduction}

In many areas of science escape over a saddle point is 
an important physical phenomenon \cite{Book1}.  
In some cases results can be derived in the 
low-noise limit \cite{LowNoise}.  
However, what is often missing is any idea of 
when the low-noise limit is applicable.  This paper addresses the 
question of how low the noise must be before the metastable 
lifetime becomes dominated by the low-noise limit results.  

Here the 
lifetime of the metastable state of the kinetic Ising model is studied.  
The square-lattice Ising model is studied using Glauber 
dynamics \cite{Glauber}, 
as well as a dynamic derived from a coupling of a 
quantum spin $1\over 2$ system 
to a 1-dimensional phonon heat bath \cite{KParkCCP,KParkUGA}.  
The simple-cubic dynamic Ising model with a Glauber dynamic 
is also simulated in the low-temperature limit.   

Although only specific cases are simulated, a reasonable hypothesis 
emerges as to the question of how low is a low enough temperature 
before the low-noise results are valid.  The hypothesis is that it depends 
on how close the applied field 
is to a value where the low-temperature prefactor has a 
discontinuity or a divergence.  As the field where the 
low-temperature prefactor has a discontinuity or divergence is 
approached, 
the low-temperature limit is seen at progressively lower temperatures.  

\section{Models and Methods}

The Hamiltonian of a spin-$1\over 2$ system can be written as 
$
{\mathcal H} = {\mathcal H}_{\rm sp} +  {\mathcal H}_{\rm bath} 
$.
The spin Hamiltonian is 
\begin{equation}
{\mathcal H}_{\rm sp} = -J \sum_{\rm nn}\sigma_i^z\sigma_j^z 
-H_z\sum_i\sigma_i^z
\label{EQHam2}
\end{equation}
where $J$ is the ferromagnetic ($J>0$) nearest-neighbor 
interaction parameter due to the exchange coupling between spins, 
$H$ is the applied external field, $\sigma^z_j$ is the 
$z$-component of the Pauli spin operator at site $j$, 
the first sum is over all nearest-neighbor (nn) 
pairs (4 nn for the square lattice and 6 for the simple-cubic lattice), 
and the second sum is over all $N$ spins.  

To simulate the quantum system given by the Hamiltonian requires 
explicit knowledge of the heat bath to which the spin system is coupled.  
In general, simulating the complete quantum-mechanical system 
to obtain the time dependence of the spin degrees of freedom is 
unnecessary.  
With the given spin Hamiltonian, ${\mathcal H}_{\rm sp}$,
the dynamic is determined by the 
generalized master equation
\cite{KParkCCP,KParkUGA,BlumBook}
\begin{eqnarray}
\! \! \frac{{\mathrm d}\rho(t)_{m^{\prime} m}}{{\mathrm d}t}&=&\frac{i}{\hbar}
[\rho(t),{\mathcal H}_{\mathrm {sp}}]_{m^{\prime} m}
+ \delta_{m^{\prime} m} \sum_{n \neq m} \rho(t)_{nn} W_{mn} 
- \gamma_{m^{\prime} m} \rho(t)_{m^{\prime} m}~,
\nonumber \\
\gamma_{m^{\prime} m}&=&\frac{W_m+W_{m^{\prime}}}{2} \;, \; \;
W_m = \sum_{k \neq m} W_{km}\;\; ,
\end{eqnarray}
where $\rho(t)$ is the time dependent density matrix of the
spin system, $m^{\prime}$, $n$, $k$, and $m$ denote the eigenstates of
${\mathcal H}_{\rm sp}$,
$\rho(t)_{m^{\prime} m}$$=$$\langle m^{\prime}| \rho(t) | m \rangle$,
and $W_{km}$ is a transition rate from the $m$-th to the $k$-th eigenstate.
For our spin Hamiltonian there are no off-diagonal components, and 
the generalized 
master equation becomes identical to the classical master equation of a 
classical spin $1\over 2$ Ising system \cite{BinderMaster} with Hamiltonian 
\begin{equation}
{\mathcal H}_{\rm Ising} = -J \sum_{\rm nn}\sigma_i\sigma_j 
-H_z\sum_i\sigma_i \; ,
\end{equation}
where $\sigma_j$$=$$\pm1$ is the classical Ising spin.  
The explicit dynamic for the system depends on the 
transition rates $W_{k m}$.  

Martin in 1977 \cite{Martin} used the 
quantum Hamiltonian of Eq.~(\ref{EQHam2}).  
He made the assumptions that each spin 
was coupled to its own fermionic heat bath, and that the 
correlation times in the heat bath are much shorter than the 
times of interest in the spin system.  He then integrated over 
all degrees of freedom of the heat bath.  He found that with 
appropriate assumptions the dynamic for the classical Ising model 
consisted of randomly choosing a spin and flipping it with 
a probability given by 
\begin{equation}
  p_{\rm G,flip} = q_{\rm G} { {\exp(\beta E_{\rm old})}
\over{\exp(\beta E_{\rm old})+\exp(\beta E_{\rm new})}} \; ,
\end{equation}
where $\beta=T^{-1}$ (with Boltzmann's constant set to unity), 
$E_{\rm new}$ is the energy of the configuration with the 
chosen spin flipped and $E_{\rm old}$ is the energy of the 
original spin configuration.  
Here $q_{\rm G}$ is an attempt frequency, related 
to the microscopic coupling between the heat bath and the spin Hamiltonian.  
Note that to insure that all probabilities are between zero and 
one requires that $0<q_{\rm G}\le 1$.  
In this paper $q_{\rm G}=1$.  
This derived dynamic corresponds the Glauber dynamic \cite{Glauber} 
of randomly choosing a spin, randomly choosing a random number 
$r$ uniformly distributed between zero and one, and flipping the 
chosen spin if $r\le p_{\rm G,flip}$.  

Recently with K.\ Park a dynamic was derived for the Ising model 
by coupling it to a phonon bath \cite{KParkCCP,KParkUGA}.  
Here we concentrate on a one-dimensional phonon bath, 
but results for phonon baths in other dimensions have also been 
obtained \cite{KParkCCP,KParkUGA}.  
Again the assumption is made that the 
correlation times in the heat bath are much shorter than the 
the times of interest in the spin system, and then the integration over 
all degrees of freedom of the heat bath is performed. 
In this case the dynamic is given by randomly choosing a spin, 
and by using a flip probability given by 
\begin{equation}
p_{\rm P,flip} = 
{q_{\rm P}} \left|
{\left(E_{\rm old}-E_{\rm new}\right) \exp(\beta E_{\rm old})
\over{\exp(\beta E_{\rm old})-\exp(\beta E_{\rm new})}}
\right|
\label{EQpflipP}
\end{equation}
for $E_{\rm old}\ne E_{\rm new}$, and zero if 
$E_{\rm old}=E_{\rm new}$.  
In Eq.~(\ref{EQpflipP}) 
the attempt frequency $q_{\rm P}$ must be chosen so as to 
make all probabilities less than one.  For the fields and temperatures 
simulated here, this is accomplished by choosing $q_{\rm P}=0.01$.  

We are most interested in measuring the lifetime of a metastable state 
for the Ising model.  We start with all spins up ($\sigma=+1$), and 
apply a static field of strength $H$ directed opposite to the spins 
(directed downward).  
The total magnetization for the model is $M=\sum_i^N\sigma_i$, 
so the magnetization starts at $M=N$.  
We measure the time $\tau$ required for the magnetization to reach 
$M=0$, since for this magnetization one has crossed the saddle point and 
is rapidly moving toward the equilibrium magnetization.  
The units for $\tau$ used are Monte Carlo steps (mcs), 
where one mcs corresponds to one attempted spin flip.  
The physical time is proportional to the time in 
units of Monte Carlo steps per spin.  
However, the 
low-temperature limit relevant here is the Single-Droplet 
regime \cite{PerSDMD} (where a single nucleating droplet causes escape 
from the metastable state).  Consequently, 
we will use mcs as our units of $\tau$.  
In the present work 
the measurement for $\tau$ is repeated for $10^3$ escapes using different 
random number sequences, and the average lifetime $\langle\tau\rangle$ 
is calculated from these escapes.  
The average lifetime in the Single-Droplet regime has the form 
\begin{equation}
\langle\tau\rangle = A \exp\left(\beta \Gamma\right) / q
\end{equation}
where $A$ is the prefactor and $\Gamma(H)$ is the 
energy of the nucleating droplet.  Note that 
the attempt frequency $q$ enters this equation in a natural way, 
so that changing $q$ will not change $A$.  
Here $q$ will stand for $q_{\rm G}$ for the Glauber dynamic or 
$q_{\rm P}$ for the phonon dynamic.  
The prefactor $A$ is a function of $T$ and $H$, and depends 
on the explicit dynamic of the system.  
In a given field, $\Gamma$ and the prefactor $A$ at zero temperature 
can be obtained from the measured lifetimes 
using a linear fit to 
\begin{equation}
T\ln\left(\langle\tau\rangle q\right) = T\ln(A) + \Gamma
.  
\end{equation}
Another way of analyzing the data \cite{SHNE02}, if the 
low-temperature value of $\Gamma$ is known, is to 
calculate an effective prefactor $A_{\rm eff}$ at any finite 
temperature, 
\begin{equation}
A_{\rm eff} = \langle\tau\rangle \> q \> \exp(-\beta\Gamma)
\; .
\label{EQAeff}
\end{equation}
In Eq.~(\ref{EQAeff}) calculating $A_{\rm eff}$ assumes that the value of 
$\Gamma$ used is the zero-temperature limit of $\Gamma$, i.e.\ the 
value of $\Gamma$ derived in the low-noise limit.  

The dynamic of randomly choosing a spin and flipping 
it or not with the decision made by comparing a random number 
$r$ with the flip probability $p_{\rm flip}$ is the physical dynamic. 
Note the time is updated whether or not the spin is flipped.   
This dynamic cannot be changed without changing the physics.  
It can be implemented in a straightforward way on serial computers.  
However, the average lifetimes can be extremely long at low temperatures, 
making simulation in the straightforward manner unfeasible.  
Although the dynamic cannot be changed, the dynamic can be implemented 
on the computer in a more efficient fashion.  One way of doing this is 
to use a rejection-free technique, by which only moves that are successful 
are implemented, and the time to make such a successful move is added to 
the current time.  For discrete models, such as the Ising model, this 
method is called the $n$-fold way, and was implemented 
in continuous time by 
Bortz, Kalos, and Lebowitz in 1975 \cite{BKT}.  
The $n$-fold way method can also be implemented in discrete time 
\cite{MANCinP}.  The $n$-fold way method can lead to exponential 
speed-ups in the algorithm.  Unfortunately, particularly at small 
field values, the $n$-fold way exponential speedup is not 
sufficient to allow low-temperature Ising simulations to progress 
in a reasonable amount of computer time.  However, the 
discrete-time version of the $n$-fold way can be further accelerated by 
realizing that the $n$-fold way algorithm uses a $1\times1$ absorbing 
Markov matrix to decide what will be 
the next spin configuration and to decide the time to 
exit from the current spin configuration.  
This Monte Carlo with absorbing Markov chain method (the MCAMC method) 
can naturally be generalized to the case of 
using $s\times s$ absorbing Markov 
Chains \cite{MAN95,MANUGA2,MANCCP01}.  
The details of the MCAMC method as well as the projective dynamics 
method, both of which are used here, can be found in a 
recent review \cite{MANreview}.  
For the simulations here, we have used $s=1$, 2, and 4 
in the absorbing Markov chains in the MCAMC 
method.  For any $s$ the average lifetime, $\langle\tau\rangle$, 
and all other averages are the same as for the straightforward 
implementation of the dynamics.  
This is because by using the MCAMC algorithm the dynamic 
has not been changed, 
the dynamic has just been implemented on the computer in a more intelligent 
fashion.  

\section{Square-lattice Ising Model}

\begin{figure}[htbp]
\includegraphics[width=.3\textwidth]{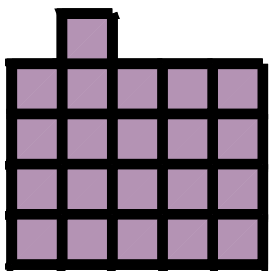}
\caption[]{The low-temperature square-lattice 
nucleating droplet with $\ell_2=5$ is shown}
\label{2dNucDrop}
\end{figure}

At very low temperatures (in the low-noise limit) the kinetic Ising model 
lifetimes are influenced by the discreteness of the lattice.  
This has allowed for exact calculation of the saddle point as well as the 
most probable route to the saddle point.  
For the square lattice with 
$\ell_2=\lfloor 2J/|H|\rfloor+1$, 
the critical droplet is a square of size $\ell_2\times\ell_2$
with one row removed and a single overturned spin on one of the 
longest sides \cite{NevSch}.  
Here $\lfloor x\rfloor$ denotes the integer part of $x$.  
Figure \ref{2dNucDrop} shows one such critical droplet for 
$\ell_2=5$, when ${2\over 5}J<|H|<{1\over 2}J$.  
The average lifetime is then given by 
$\langle\tau\rangle q=A_2\exp\left(\beta\Gamma_2\right)$
with 
$\Gamma_2=8J\ell_2-2|H|(\ell_2^2-\ell_2+1)$.  
This is valid for low temperatures and for $|H|<4J$ \cite{NevSch}.  

\subsection{Glauber Dynamics}

\begin{figure}[htbp]
\includegraphics[width=.6\textwidth]{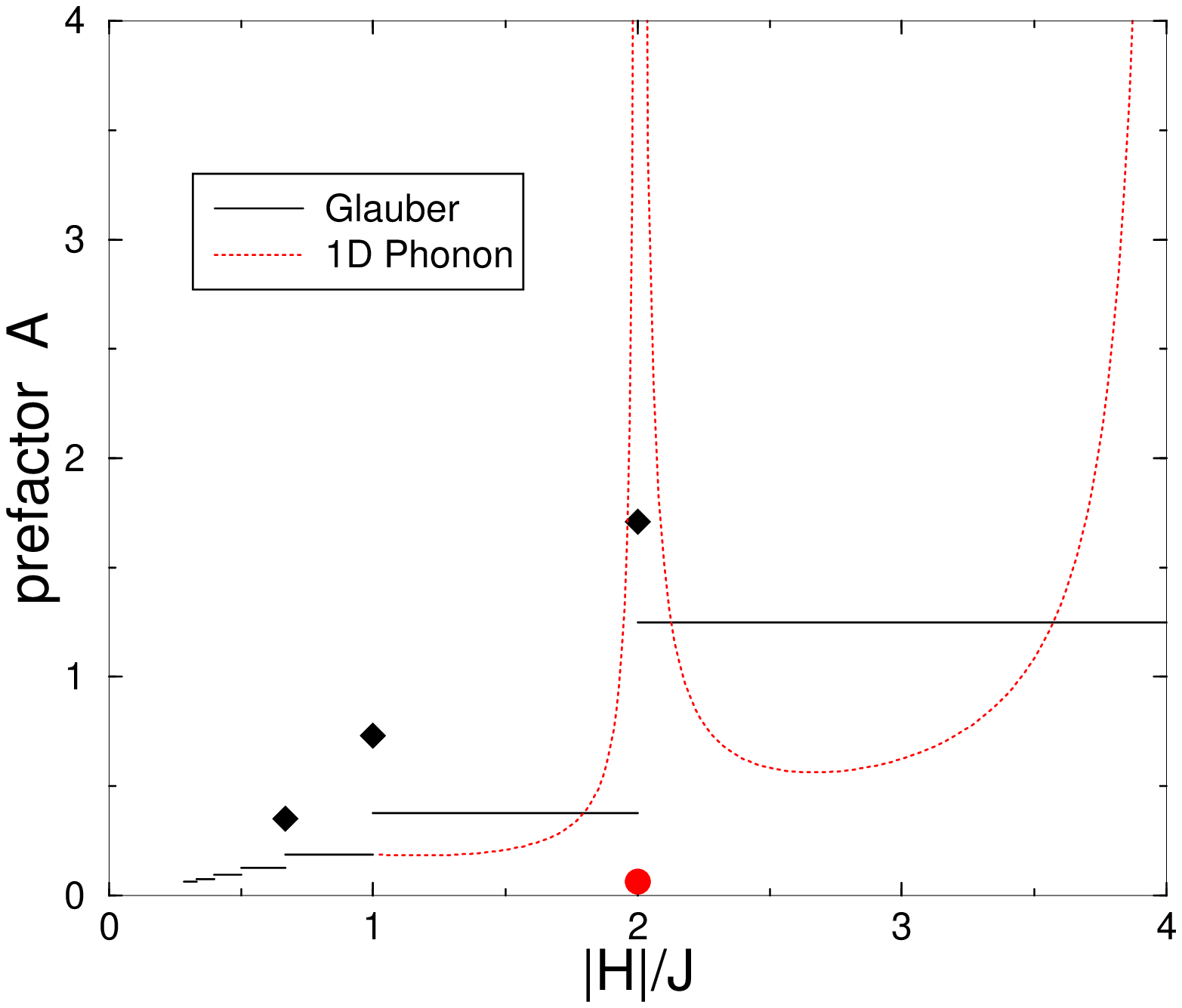}
\caption[]{The low-temperature square-lattice prefactors 
are shown vs.\ $|H|$.  The solid lines 
and diamonds are the 
exact low-temperature results for the prefactors for the 
Glauber dynamic.  
The dashed curve and filled circle are the low-temperature 
prefactors for the Ising model with each spin connected to a 
$d=1$ phonon bath.  
The curves and lines are derived analytically, while the 
symbols are from low-temperature MCAMC simulations. 
The prefactors are shown only for large values of $|H|$.  
Only three prefactors for the Glauber dynamic have been obtained 
at the discontinuity. Only prefactors for $|H|\ge J$ are shown 
for the phonon dynamic}
\label{3dphonon1}
\end{figure}

The prefactor for the square lattice 
with the Glauber dynamic was determined from absorbing Markov chain 
calculations to be $A_2={5\over 4}$ for $\ell_2=1$ and 
$A_2={3\over 8}$ for $\ell_2=2$ \cite{MANUGA2}.  Recently the 
prefactor has been found to be 
$A_2=3/[8(\ell_2-1)]$ for $\ell_2>1$ \cite{RecPref}.  
At low enough temperature these results should hold, except when 
$2J/|H|$ is an integer.  
Figure \ref{3dphonon1} shows these prefactors for many values of 
$H$.  Note that there is a discontinuity in the prefactor (but not 
in $\Gamma_2$) when $2J/|H|$ is an integer.  

\begin{figure}[htbp]
\includegraphics[width=.49\textwidth]{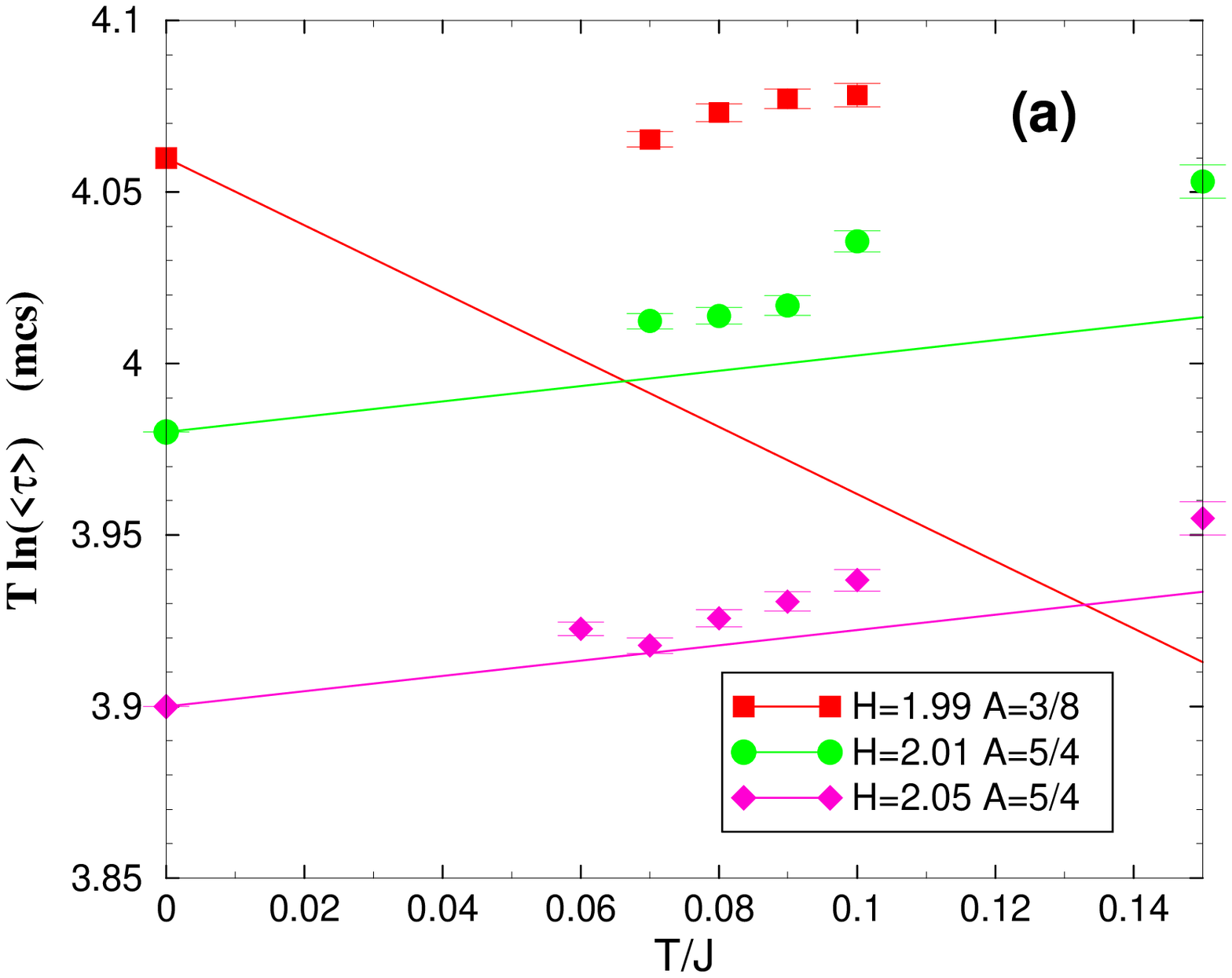}
\includegraphics[width=.49\textwidth]{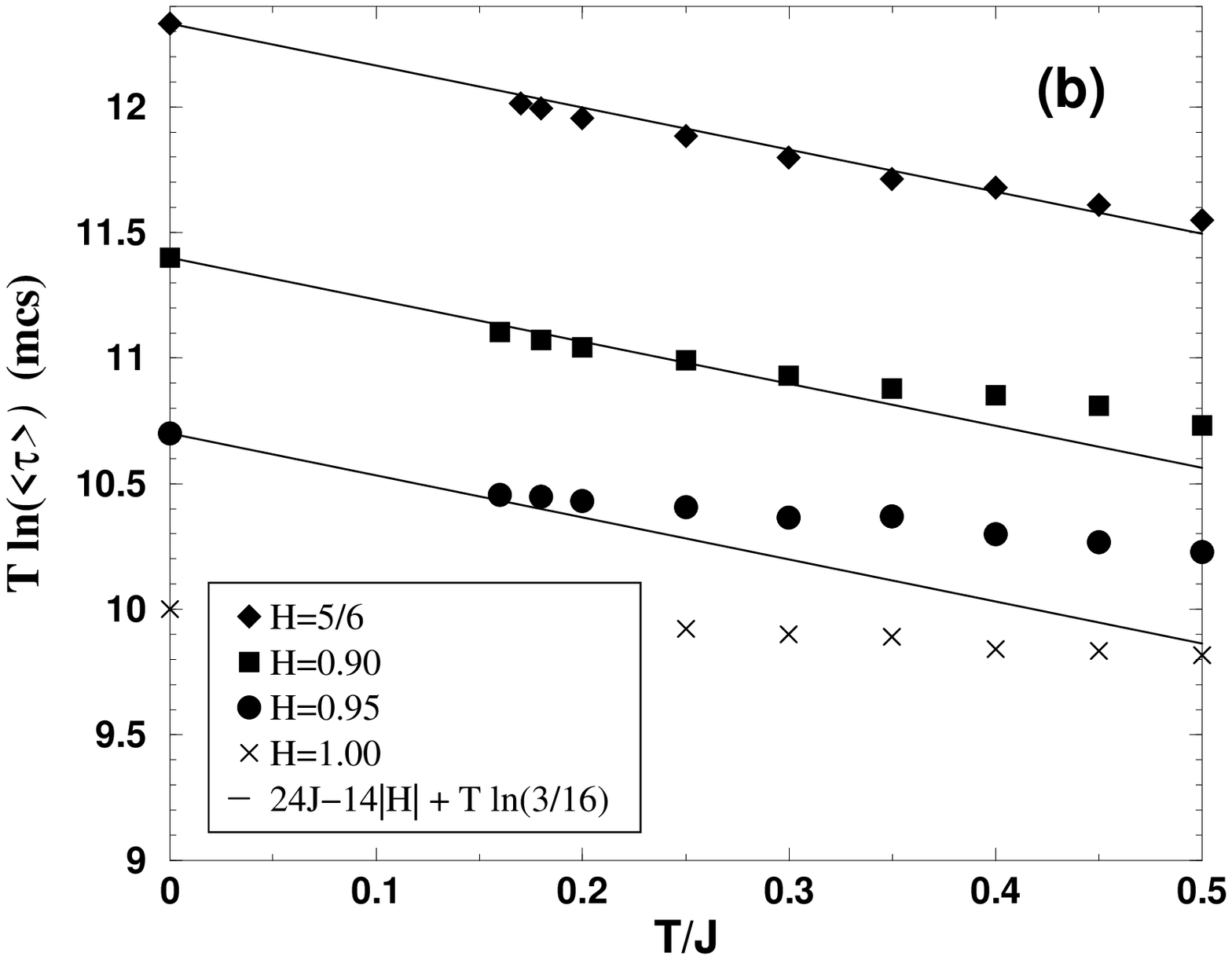}
\caption[]{The low-temperature results for the square-lattice Ising 
model as functions of $T$.  The solid lines are the 
exact low-temperature results, including the prefactors.  
(a) Three different fields are shown.  
Note that the lines cross, which would mean a non-monotonic 
behavior of the lifetime in this region.
(b) 
Results for $\ell_2=3$ are shown 
for several values of $|H|$.  Note that as the discontinuity 
at $|H|$$=$$J$ is 
approached, the temperature required before the 
low-temperature results are seen decreases}
\label{2dlowtnonmono}
\end{figure}

The exact values of $\Gamma_2$ and $A_2$ are known, so the 
MCAMC data at finite temperatures can be compared with these 
predictions.  Figure \ref{2dlowtnonmono}(a) shows that 
near $|H|=2J$ the predicted lines of $T\ln(\langle\tau\rangle)$
vs.\ $T$ cross.  
If the measured lifetimes were to follow these 
expected curves, it would mean that for $T>0.07J$ the lifetime at 
a field of $|H|=1.99J$ would be smaller than for a field of 
$|H|=2.01J$.  In other words, there would be regions where the lifetime 
decreases as the field decreases.  
As seen in Fig.\ \ref{2dlowtnonmono} this does not actually occur.  
Rather, the low-temperature predictions only agree with the data at 
lower and lower temperatures as the value of $|H|$ approaches the 
value $2J/|H|$ where the prefactor has a discontinuity.  
This is also demonstrated in Fig.\ \ref{2dlowtnonmono}(b).  
Note that these 
MCAMC results using a Glauber dynamic are for $10^3$ escapes from the 
metastable state for a $24\times24$ square lattice Ising system.  
The error estimates are from the second moment of $\tau$ obtained 
from the $10^3$ escapes.  These error estimates should be viewed 
as approximate, since the measured second moment most likely 
deviates substantially 
from the exact one when only $10^3$ escapes are used.  

\begin{figure}[htbp]
\includegraphics[width=.5\textwidth]{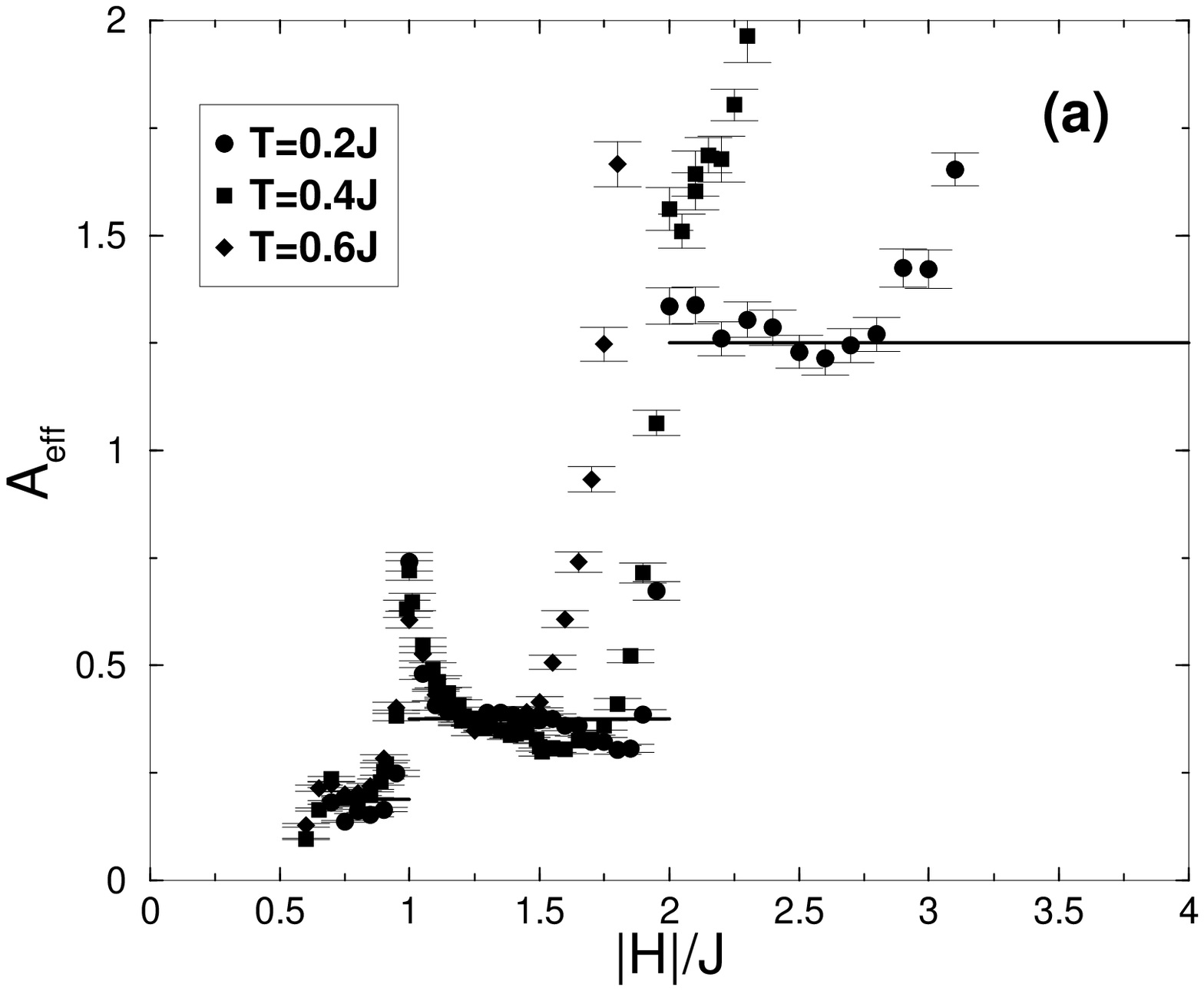}
\includegraphics[width=.5\textwidth]{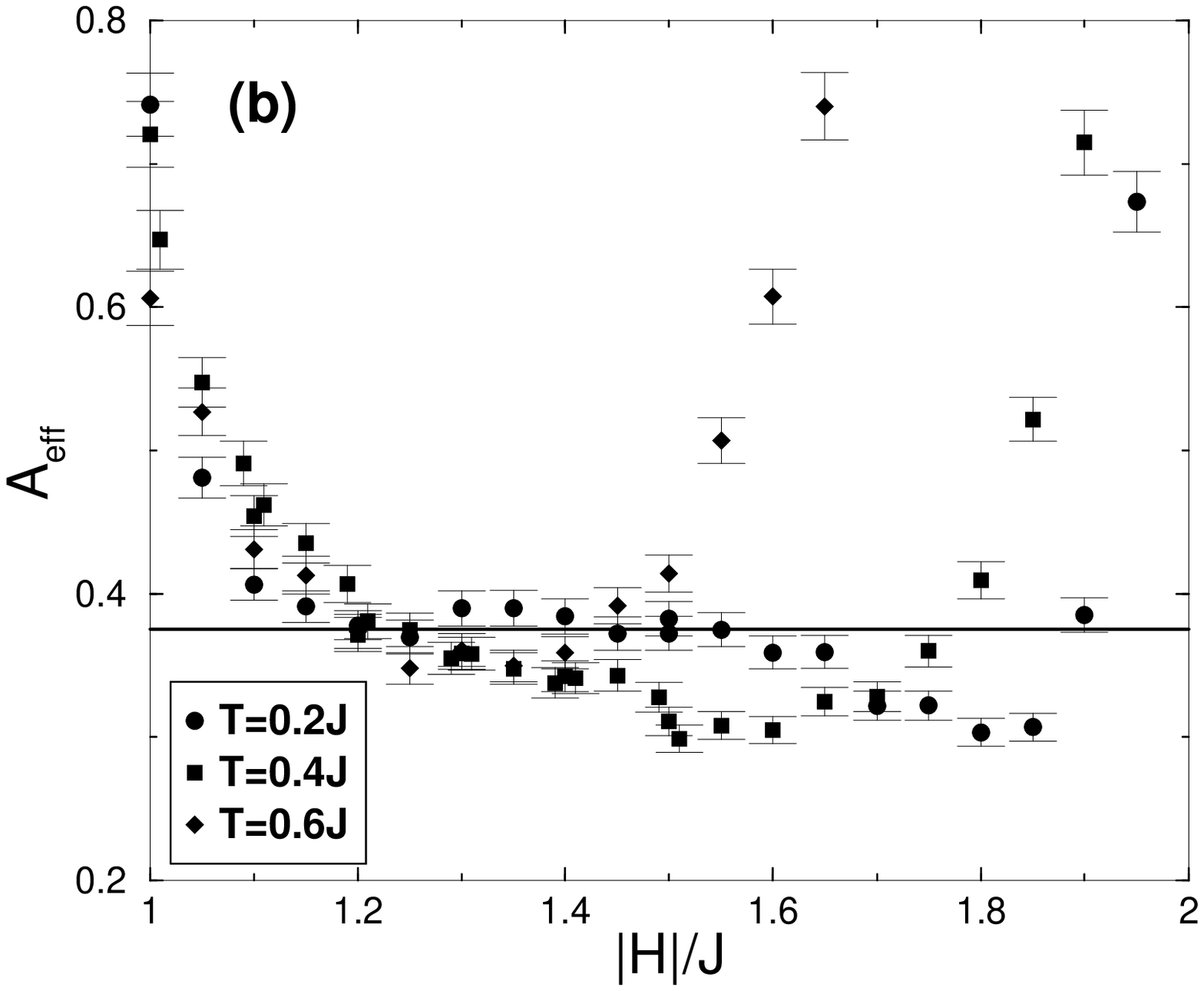}
\caption[]{
The low-temperature results for the effective prefactor 
in the Single Droplet regime 
are shown as a function of $|H|$.  The solid horizontal lines are the 
exact low-temperature prefactors.  Three different temperatures are shown.  
Note that the deviation from the expected results at larger $|H|$ values 
can be mainly attributed to the cross-over between the 
Single and Multi-Droplet regimes. (a) For ${1\over 2}J\le |H|\le 4J$.  
(b) Only for $J\le |H|\le 2J$, where $\ell_2$$=$$2$}
\label{2dlowtl}
\end{figure}

Figure \ref{2dlowtl} shows the results for $A_{\rm eff}$ obtained 
at three different temperatures.  
Only MCAMC data points in the Single Droplet regime (as defined here 
by the ratio of the second moment of $\tau$ to $\langle\tau\rangle$ 
being greater than $1\over 2$ \cite{PerSDMD}) are shown.  
It is seen that the exact low-temperature prefactor is approached 
at finite temperature more quickly for values of $H$ that are far 
from the values where $A$ has a discontinuity.  
Note that in Fig.\ \ref{2dlowtl} the higher $|H|$ values start a 
cross-over toward the multi-droplet regime \cite{PerSDMD}, 
and so should not be used in the comparison with the low-temperature 
prefactors.  This cross-over depends on the system size as well 
as the temperature \cite{PerSDMD}.  
Figure \ref{2dlowtl}(b) shows a portion of the data in 
Fig.\ \ref{2dlowtl}(a) for $\ell_2=2$.  
Again note that near a value of $H$ where there is a 
discontinuity (here $|H|=J$ and $2J$) the effective prefactor 
at finite temperature approaches the 
exact low-temperature prefactor extremely slowly.  

A recent preprint \cite{SHNE02} has performed a similar analysis of 
the prefactors for the square-lattice Ising dynamic using a generalized 
Becker-D{\"o}ring approach.  The authors find qualitatively similar results 
to those shown in Fig.\ \ref{2dlowtl}.  However, unlike our results, 
their results do not approach the predicted low-temperature prefactors.  
They argue that this might be due to a 
difference between discrete and continuous times in the Glauber 
simulations.  
However, for the temperatures and field values simulated here we did 
not observe in our simulations that the prefactor value would depend on 
whether the time in the Glauber simulation is continuous or discrete.  
In the single-droplet regime relevant at low 
temperatures, the average lifetimes are extremely long, and 
consequently the average lifetimes should not depend in a significant 
way on whether the time in the simulation is continuous or discrete.  
The dynamic used in ref.~\cite{SHNE02} is not identical to the Glauber 
dynamic, and this difference between the two dynamics might 
account for the difference in the prefactors.  

\subsection{Phonon Dynamics}

\begin{figure}[htbp]
\includegraphics[width=.6\textwidth]{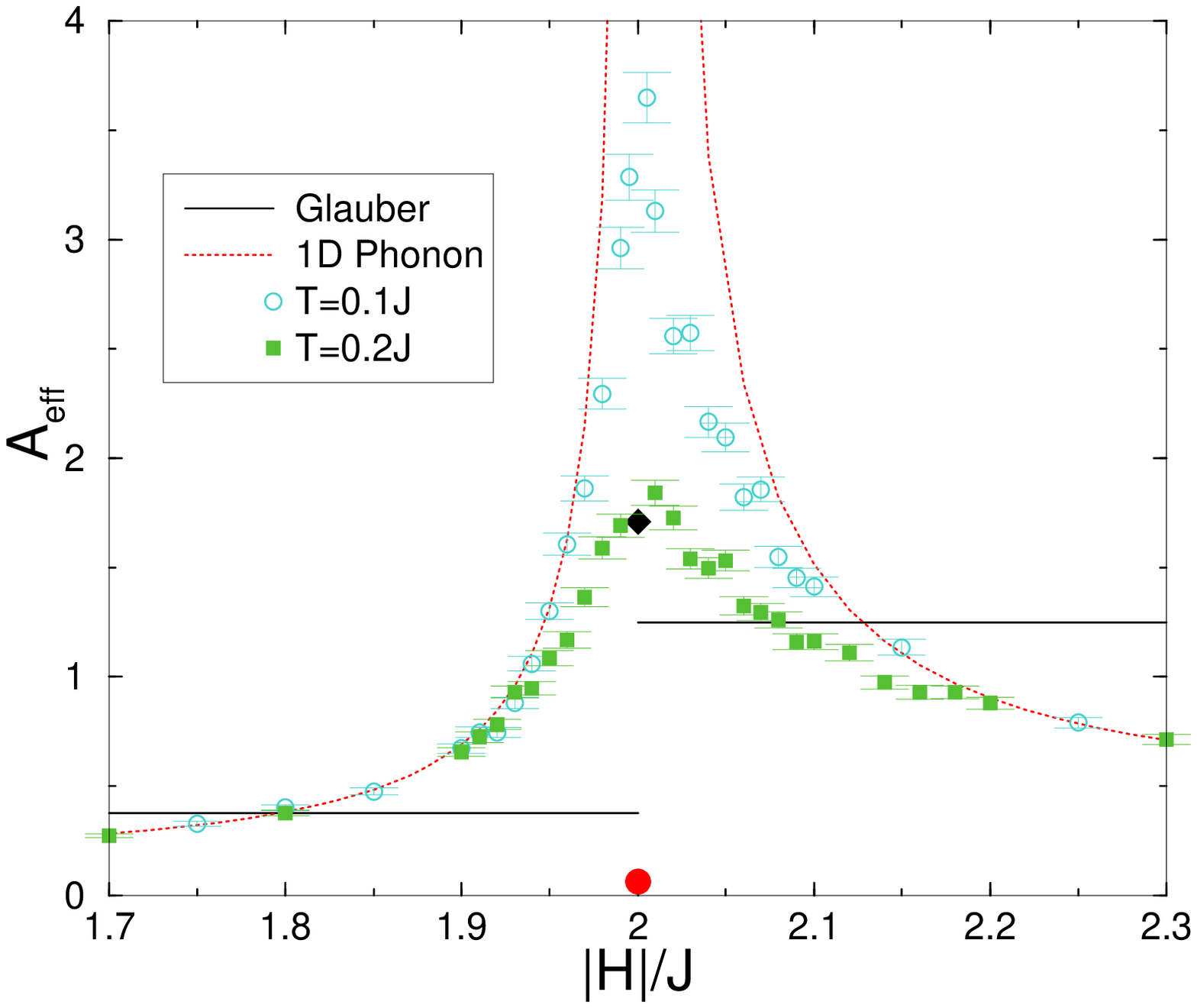}
\caption[]{The low-temperature results for the prefactors 
vs.\ $|H|$.  The solid lines 
and diamond are the 
exact low-temperature results for the prefactors for the 
Glauber dynamic.  
The dashed curve and filled circle are the low-temperature 
prefactor for the Ising model with each spin connected to a 
$d=1$ phonon bath.  
The curves and lines are derived analytically, while the 
symbols are from low-temperature Monte Carlo simulations.
The effective prefactor from MCAMC data near $H=2J$ is shown 
for $10^3$ escapes from the metastable state on 
a $32\times32$ lattice at $T=0.1J$ and $T=0.2J$.  Note that 
the low-temperature results are approached very slowly near the 
discontinuity}
\label{1dphononNear2}
\end{figure}

For the square-lattice Ising model coupled to a one-dimensional 
phonon bath Fig.\ \ref{3dphonon1} shows the exact low-temperature 
prefactor \cite{KParkCCP} for $|H|>J$.  This is given by the equation 
\begin{equation}
A={{4(2|H|-4J)+(8J-2|H|)}\over{4(2|H|-4J)(8J-2|H|)}}
\end{equation} 
for $2J<|H|<4J$ and 
\begin{equation}
A={{|H|+2(2J-|H|)}\over{2^4 |H|(2J-|H|)}}
\end{equation} 
for $J<|H|<2J$ \cite{KParkCCP}.  
The same nucleating droplets (except at $|H|=2J$) 
are responsible for the decay of the metastable state for this dynamic 
as for the Glauber dynamic, so the same value of $\Gamma_2$ is 
found for both dynamics (except at $|H|=2J$).  
Note that for this dynamic the prefactor is not constant when 
$\ell_2$ is constant, and there is a divergence in the 
prefactor when $|H|=2J$ or $|H|=4J$.  The divergence in the 
prefactor at $|H|=2J$ is due to the nucleating droplet at this 
field being two next-nearest neighbor overturned spins that has 
$\Gamma=8J$ rather than the value $\Gamma_2=4J$ for the Glauber dynamic.  
Recently, such a divergence in the prefactor has also been found 
in a model with continuous degrees of freedom \cite{MaiStePRL}.  
Figure \ref{1dphononNear2} shows values of $A_{\rm eff}$ 
(from finite-temperature MCAMC) 
at two different temperatures as they approach the exact low-temperature 
results.  Note that the divergence in the prefactor is only seen 
at lower temperatures.  

\section{Simple-cubic lattice Ising Model}

\begin{figure}[htbp]
\includegraphics[width=.4\textwidth]{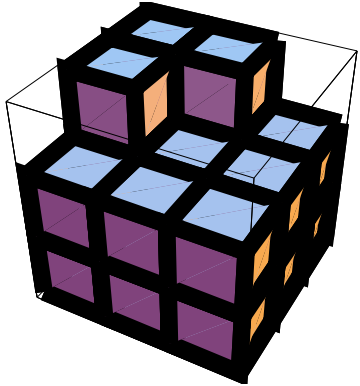}
\caption[]{A low-temperature 
nucleating droplet with $\ell_3=3$ and $\ell_2=2$.  
It consists of a 
cube of overturned spins with sides of length $\ell_3$, with the spins from 
one face removed and a nucleating droplet with $\ell_2$ 
placed on the removed face}
\label{3dNucDrop}
\end{figure}

For the kinetic Ising model with the Glauber dynamic 
on a simple cubic lattice the average lifetime in the 
low-temperature limit (for fixed $H$ and system size) is given by 
$
\langle \tau\rangle = A_3\exp(\beta\Gamma_3) / q
$
with 
\begin{equation}
\Gamma_3 = 12J\ell_3^2-8J\ell_3-2|H|\ell_3^2(\ell_3-1)+ \Gamma_2
\label{EqGamma3}
\end{equation}
with 
$\ell_3=\lfloor4J/|H|\rfloor+1$ \cite{RecPref,CFQ1,CFQ2} 
and $\Gamma_2$ the square-lattice nucleating droplet energy defined above.  
Equation~(\ref{EqGamma3}) should be valid when $|H|\le4J$ and 
$2J/|H|$ is not an integer.  
The number of spins in the nucleating droplet 
is $n_c=\ell_3^2(\ell_3-1)+(\ell_2^2-\ell_2+1)$.  
The nucleating droplet is a cube with $\ell_3$ overturned spins per side, 
with one face removed and the appropriate square-lattice nucleating 
droplet placed where the face was removed.  
Figure \ref{3dNucDrop} shows a nucleating droplet for 
$J\le|H|\le2J$, so $\ell_3=3$ and $\ell_2=2$, and the nucleating 
droplet has $n_c=21$.  For this value of 
$H$ the square-lattice nucleating droplet consists of three overturned 
spins in an L-shape.  
The prefactor for the simple cubic lattice with the Glauber dynamic 
has recently been found to be 
$A_3=\left[16(\ell_3-\ell_2+1)(\ell_2-1)\right]^{-1}$ 
for $\ell_3\ge3$ \cite{RecPref}.  
The low-temperature prefactor for 
$\ell_3=2$ (so $2J<|H|<4J$) has not yet been derived, being a 
straightforward but lengthy calculation.  
The low-temperature prefactor for $\ell_3=1$ (so $4J<|H|<6J$)
is easily derived to be $A_3=7/6$.  

\begin{figure}[htbp]
\includegraphics[width=.5\textwidth]{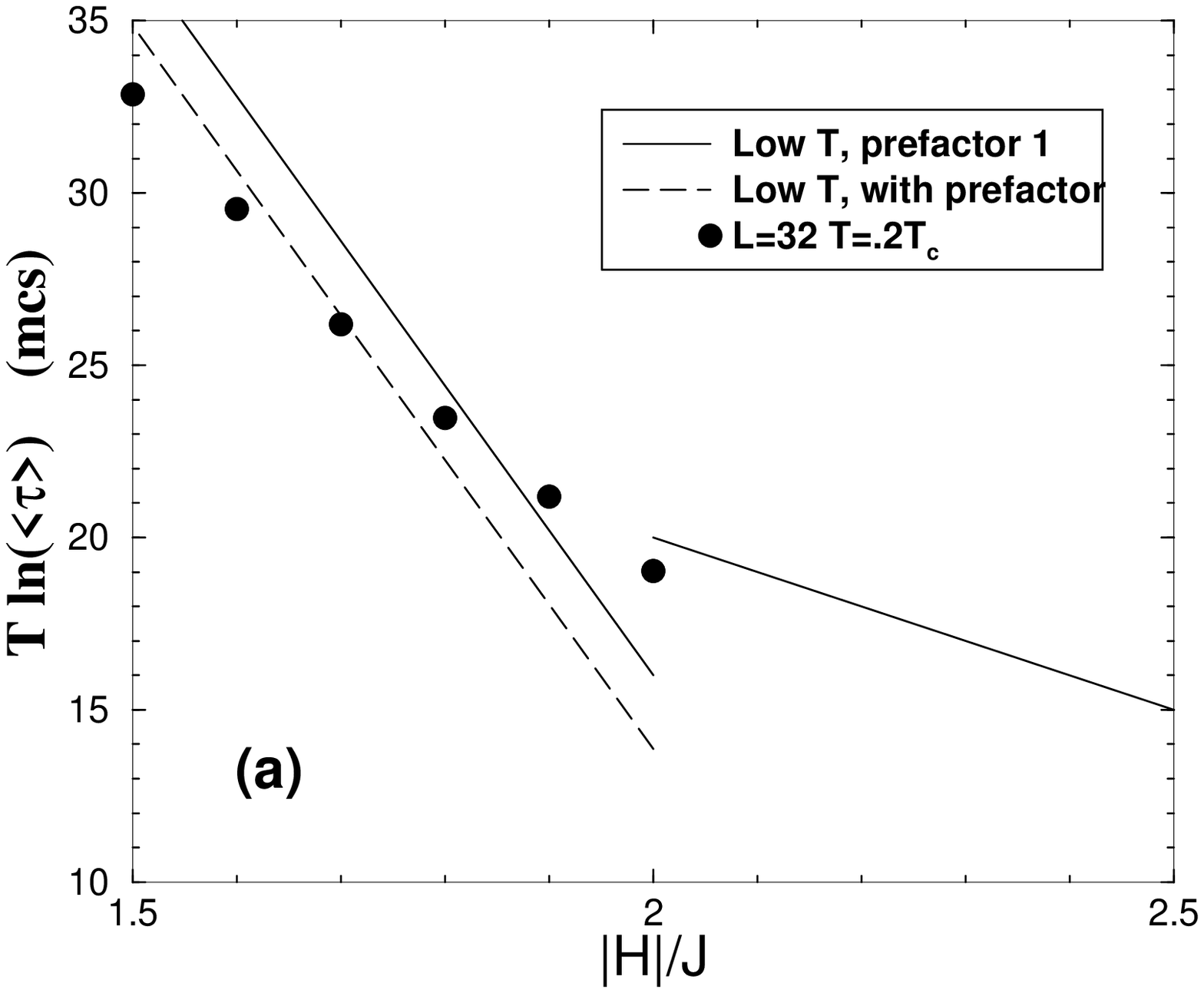}
\includegraphics[width=.5\textwidth]{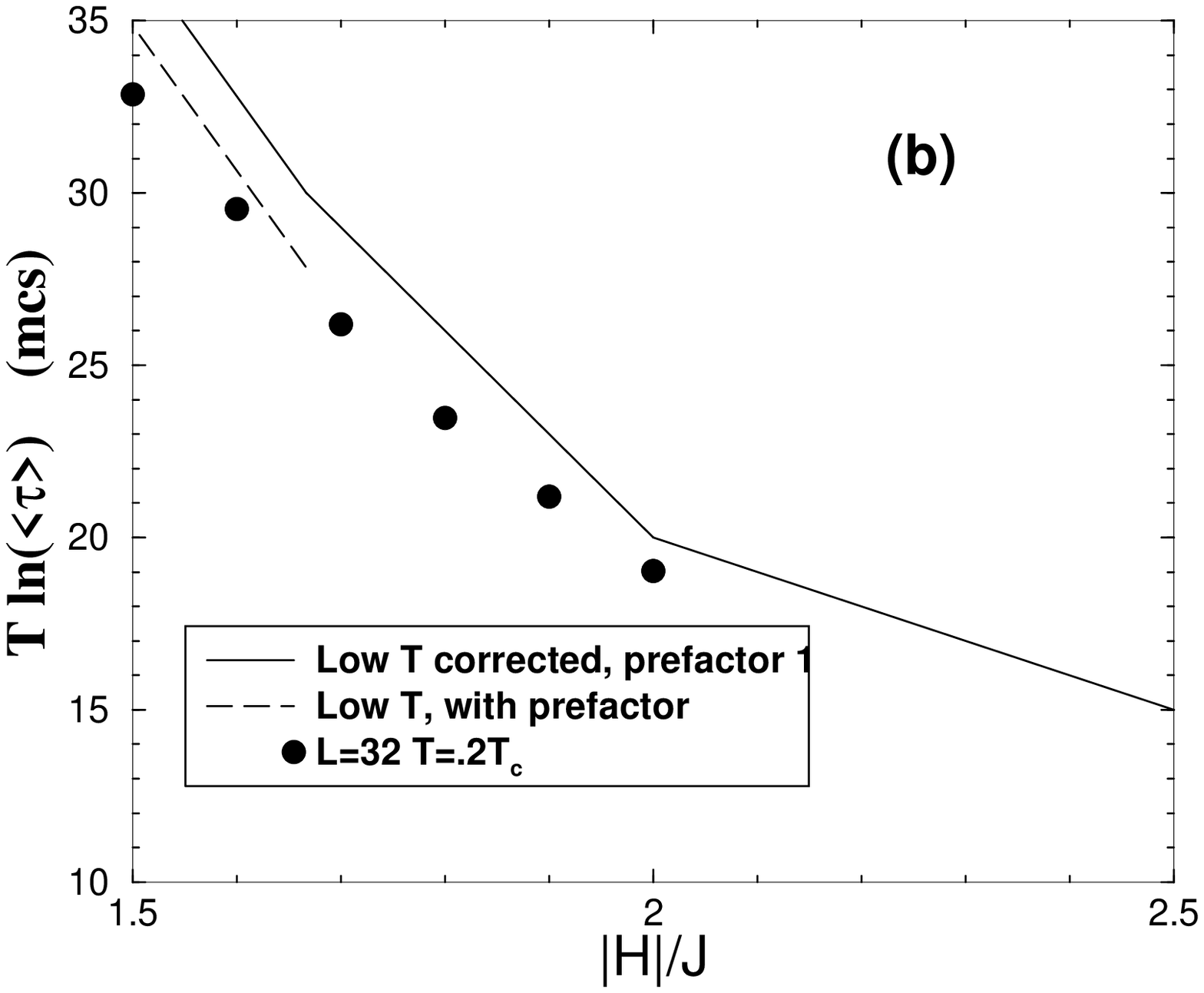}
\caption[]{The low-temperature results for the 
simple cubic lattice 
vs.\ $|H|$.  
The dashed line includes 
known exact prefactors.  The symbols are projective dynamics simulation 
results on a $32^3$ lattice at $T=0.2T_{\rm c}$.
(a) The solid lines are the exact low-temperature 
results \protect\cite{RecPref,CFQ1,CFQ2}.  
(b) The solid lines are the exact, corrected, 
low-temperature results.  The dashed line shows the known 
exact prefactors, which in this figure are only known for 
$|H|<{5\over 3}J$.  
Note that now the exponential portion of the low-temperature 
prediction is continuous}
\label{3dlowt}
\end{figure}

Figure \ref{3dlowt}(a) shows the expected low-temperature result 
near $|H|=2J$ for the simple cubic Ising model as a function of $|H|$. 
Projective dynamics \cite{MANreview,MiroPRL,Granada} 
simulation results are also shown for $T=0.2T_{\rm c}$.  
The critical temperature is $T_{\rm c}\approx4.51J$.  
If the prefactor were unity, the expected low-temperature result 
(solid line) would be independent of temperature.  For other prefactors, 
the dashed line is adjusted so the result with the prefactor depends on $T$.  
The simulation data are for a $32^3$ Ising system using 
a projective dynamic method \cite{MANreview,MiroPRL}.  
Note that away from a discontinuity the simulation data 
fall quite nicely on the expected low-temperature results including 
the predicted prefactor.  However, they do not follow the expected 
results near the discontinuities (where $\ell_3$ changes).   Shown 
here is only the discontinuity near $|H|=2J$.  This is similar to the 
result found in the square-lattice model.  Namely, how low in temperature 
the simulation must be performed before the low-temperature result is 
valid depends strongly on how close one is to a discontinuity in the 
low-temperature prefactor.  

Note that, as seen in Eq.~(\ref{EqGamma3}) and in Fig.\ \ref{3dlowt}(a), 
there is a discontinuity in $\Gamma_3$ whenever $\ell_3$ changes.  For 
all values of $H$ where $\ell_3$ changes 
the discontinuity is such that 
$\Gamma_3$ decreases by $4J$.  If the data were to follow these 
low-temperature results, as $|H|$ decreases through one of these 
discontinuities the average lifetime would decrease.  
This would occur not just because of a discontinuity in the prefactor, 
as in the square-lattice Ising model, but also because of a 
discontinuity in the exponential (in $\Gamma_3$).  

\begin{figure}[htbp]
\includegraphics[width=.6\textwidth]{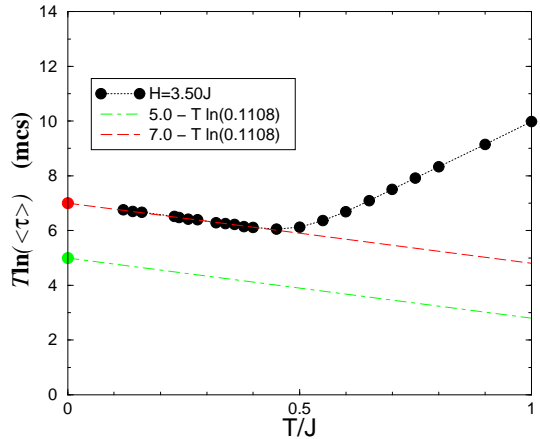}
\caption[]{The low-temperature results for the 
simple cubic lattice as a function of $T$ for $|H|=3.5J$.  
The lines have a slope of the best fit to the low-temperature 
data.  The results at $T=0$ are the exact low-temperature results, 
both the uncorrected ones (lower point) and the corrected 
ones (upper point)}
\label{3dlowH3p5}
\end{figure}

Figure \ref{3dlowH3p5} shows MCAMC simulation results using the 
Glauber dynamic for the simple-cubic Ising model at $|H|=3.5J$.  
For this value of $H$ one has $\ell_3=2$ and $\ell_2=1$ so 
the nucleating droplet is predicted to have $n_c=5$ overturned spins, 
and the predicted value of $\Gamma_3=5J$.  The low-temperature 
prefactor is not known for this value of $H$.  
As seen in Fig.\ \ref{3dlowH3p5}, the low-temperature MCAMC data 
do not tend toward $\Gamma_3=5J$, but could reasonably tend toward 
$\Gamma_3=7J$.  
The published low-temperature predictions \cite{RecPref,CFQ1,CFQ2} 
{\it do not agree\/} with our MCAMC results.  
The reason for this disagreement is that the 
previously reported low-temperature 
predictions are incomplete: they give 
accurate values for $\Gamma_3$ from Eq.~(\ref{EqGamma3}) 
for some but not all values of $H$.  This is because the 
most probable path of escape from the metastable state has a droplet 
with higher energy than predicted by $\Gamma_3$ from 
Eq.~(\ref{EqGamma3}) for some values of $H$.  
For $3J\le|H|\le4J$ this higher energy saddle has an energy of 
$28J-6|H|$, that corresponds to three overturned spins in an L-shape, 
and so has $n_c=3$.  For all $|H|<4J$ this droplet lies along the 
most probable path to the saddle point, but it is the highest 
energy point along this path only for $3J\le|H|\le4J$, and hence is 
then the critical nucleating droplet.  
This scenario predicts that for $|H|=3.5J$ one should have $\Gamma_3=7J$, 
in agreement with the MCAMC results shown in Fig.\ \ref{3dlowH3p5}.  

Similar results hold near other values of $H$ 
where $\ell_3$ changes.  For example, near  
$|H|=2J$ the low-temperature predictions \cite{RecPref,CFQ1,CFQ2} 
are that 
for $n_c=5 $ ($2J<|H|<4J$) one has $\Gamma_3= 40J-10|H|$ and 
for $n_c=21$ ($ J<|H|<2J$) one has $\Gamma_3=100J-42|H|$ 
(Fig.\ \ref{3dNucDrop}).  
However, there is a droplet with 15 overturned spins 
consisting of $2$$\times$$2$$\times$$3$ overturned spins with a 
$\ell_2=2$ droplet of 
overturned spins on one of the $2$$\times$$3$ faces.  
This 15-spin droplet has $\Gamma= 80J-30|H|$.  
This is the nucleating droplet for 
values of $H$ when this $\Gamma$ is larger than the 
predicted values of $\Gamma_3$ for $n_c=5$ or $n_c=21$.  
This occurs for ${5\over 3}J<|H|<2J$.  
At $|H|=2J$ the droplets with 5 and with 15 overturned spins have 
the same energy, while at 
$|H|={5\over 3}J$ the droplets with 15 and with 21 overturned 
spins have the same energy.  
The correct predicted values of 
$\Gamma_3(H)$ are hence continuous 
when this droplet with 15 overturned spins is included.  
This is shown in Fig.\ \ref{3dlowt}(b).  
In general, one can show that all the 
predicted \cite{RecPref,CFQ1,CFQ2} discontinuities 
in $\Gamma_3$ where $\ell_3$ changes vanish.  
After this study was completed another paper 
describing the low-temperature properties of the simple cubic 
dynamic Ising model \cite{EJP96} was brought to the author's attention.  
In ref.~\cite{EJP96} it is shown 
that the critical droplet size is 
$n_c=\ell_3(\ell_3-\delta)(\ell_3-1)+(\ell_2^2-\ell_2+1)$ 
where $\delta$$=$$0$ 
unless $4J+\sqrt{16J^2+|H|^2}<|H|(2\ell_3-1)$
in which case $\delta$$=$$1$.
For the simple cubic lattice, the critical value of 
$\Gamma_3(H)$ is thus continuous for all $H$, 
just as for the square lattice.  

\section{Summary and Conclusions}

From long-time simulations of kinetic Ising models, the question of 
when the low-noise limit results are approached by 
finite-temperature simulations has been addressed.  
In particular, we have studied the limit where the system size 
and $H$ are fixed, while the temperature is lowered.  
There are three main features that emerge:
\begin{itemize}
\item The square-lattice and simple-cubic-lattice 
low-temperature predictions have the form 
$\langle\tau\rangle=A\exp(\beta\Gamma)$ with 
$\Gamma(H)$ continuous almost everywhere, but with $A(H)$ discontinuous.  
\item The finite-temperature simulation results must be at a 
lower temperature to see the low-temperature predictions 
if the applied field is near a value where the 
low-temperature prefactor has a discontinuity or a divergence.  
\item The finite-temperature results have average lifetimes 
$\langle\tau\rangle$ which are decreasing functions of $|H|$.  
This is true even if the low-temperature results are not always 
decreasing as $|H|$ increases.  
\end{itemize}

We have shown that these features are true for simulations 
in the square-lattice and simple-cubic-lattice kinetic Ising model 
with various dynamics.  
Whether or not these features of $\langle\tau\rangle$ 
can be shown to be true 
in general is an interesting question for future research.  

\noindent{\textbf{Acknowledgements}} \\
Special thanks to M.\ Kolesik for allowing inclusion of 
unpublished projective dynamic data.  
Thanks to P.A.\ Rikvold and K.\ Park for many useful discussions.    
Partially funded by NSF DMR-0120310.  
Supercomputer time provided by the DOE through NERSC.

%

\end{document}